\documentclass[prl,superscriptaddress,showpacs,twocolumn,email]{revtex4}
\usepackage{graphicx}
\usepackage{amsmath}
\usepackage{amsfonts}
\usepackage{color}
\usepackage{amssymb}
\begin{document}
\author{Andreas Flesch}
\affiliation{
Institute for Advanced Simulation and JARA, Forschungszentrum J\"ulich, 
52425 J\"ulich, Germany}
\author{Guoren Zhang}
\affiliation{
Institute for Advanced Simulation and JARA, Forschungszentrum J\"ulich, 
52425 J\"ulich, Germany}
\author{Erik Koch}
\affiliation{German Research School for Simulation Sciences, 52425 J\"ulich, Germany}
\author{Eva Pavarini}
\email{e.pavarini@fz-juelich.de}
\affiliation{
Institute for Advanced Simulation and JARA, Forschungszentrum J\"ulich, 
52425 J\"ulich, Germany}
\title{Orbital-order melting in rare-earth manganites: the role of super-exchange}

\begin{abstract}
We study the mechanism of orbital-order melting observed at temperature $T_{\rm OO}$ in the series of 
rare-earth manganites. We find that many-body super-exchange yields
a transition-temperature $T_{\rm KK}$ that decreases with decreasing rare-earth radius,
and increases with pressure, opposite to the experimental  $T_{\rm  OO}$. 
We show that the tetragonal crystal-field splitting reduces $T_{\rm KK}$ further increasing the discrepancies with experiments.
This proves that super-exchange effects, although very efficient, in the light of the
experimentally observed trends, play a minor role for the melting of orbital ordering in rare-earth manganites. 
\end{abstract}
\maketitle

The role of orbital degrees of freedom \cite{KK} in the physics of LaMnO$_3$, and in particular the co-operative
Jahn-Teller transition, has been debated since long \cite{KK,phonons,reviews,Ku,recent}.  
Ab-initio LDA+$U$ calculations show that Coulomb repulsion effects 
are key to understanding the orbitally-ordered antiferro-magnetic ground state \cite{Ku}.
Super-exchange alone, however, is not sufficient \cite{OO} to explain the presence of co-operative Jahn-Teller 
distortions in nanoclusters up to  $T\sim$1150~K  \cite{nanoclusters,persistence}  (orbitally disordered phase).
Still, superexchange effects are rather large:  $T_{\rm KK}$, the temperature at which superexchange alone would drive the transition,
is remarkably close to  $T_{\rm OO}$, the temperature at which the {\em co-operative} Jahn-Teller distortion disappears
in resonant X-ray and neutron scattering \cite{ordertodisorder}.
This fact could indicate that super-exchange, although insufficient to explain the persistence
of Jahn-Teller distortions in the orbitally disordered phase, plays a major role in the 
orbital order-to-disorder transition (orbital order melting) observed at $T_{\rm OO}$. 
Here we resolve this issue.

\begin{figure}
\center
\rotatebox {90}{\includegraphics [width=0.26\textwidth]{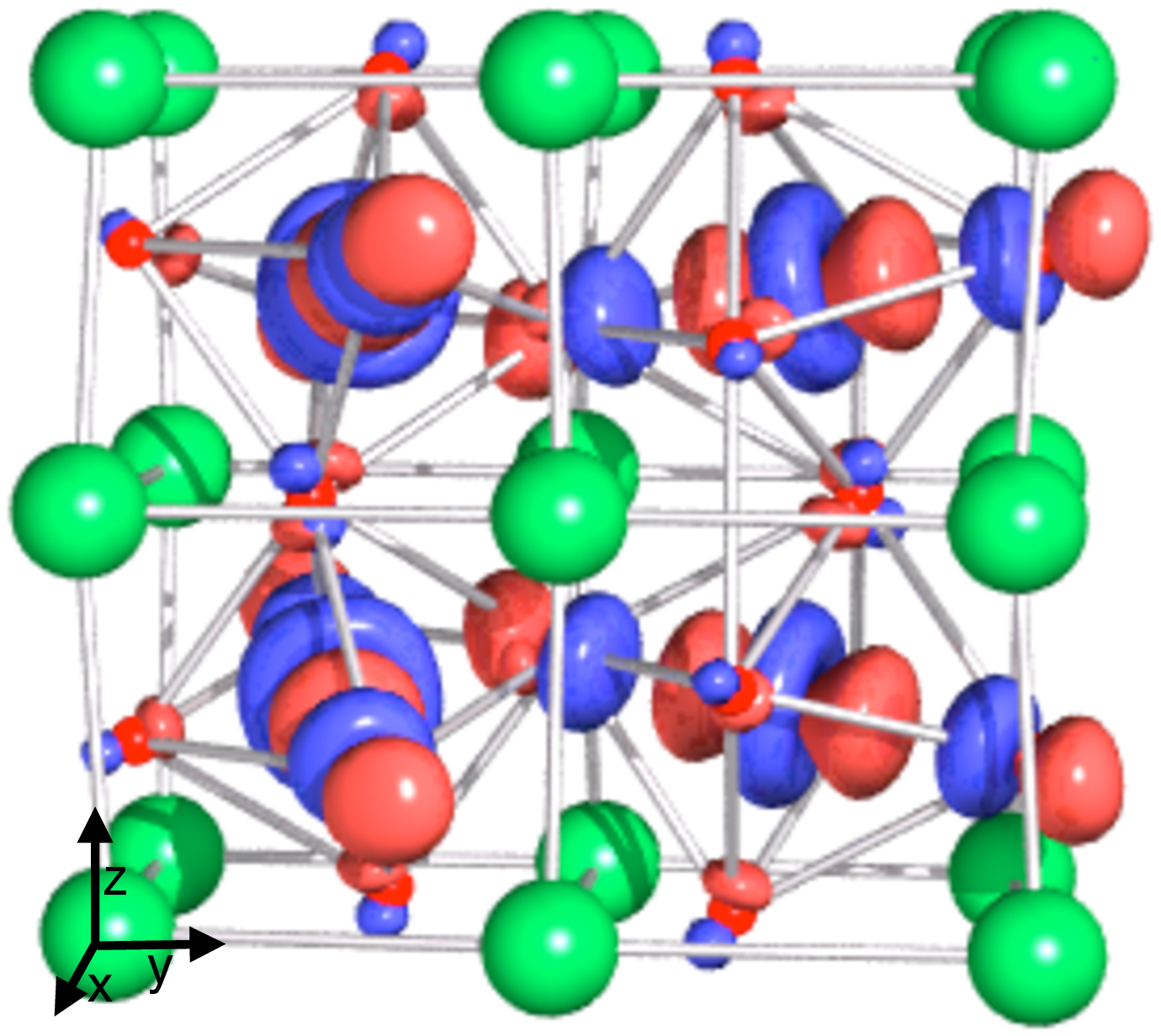}}
\caption{(Color online) \label{fig1}
Orbital-order  in TbMnO$_3$, as obtained by LDA+DMFT calculations.
The pseudo-cubic axes pointing along Mn-Mn bonds are shown in the left corner.}
\end{figure}

Remarkably, orbital-order melting has been observed \cite{Zhou2006,dabrowski,maris} in the full series of 
orthorhombic rare-earth (RE) manganites REMnO$_3$. These systems are perovskites (Fig.~\ref{fig1}) with 
electronic configuration Mn 3$d^4$ ($t_{2g}^3 e_g^1$). 
In the co-operative Jahn-Teller 
phase ($T<T_{\rm OO}$), the MnO$_6$ octahedra are tilted and rotated, 
and exhibit a sizable Jahn-Teller distortion with long and short MnO bonds
antiferro-ordered in the {$xy$} plane, and ferro-ordered along {\bf z}  (Fig.~\ref{fig1}). 
Neutron and X-ray diffraction data show that $T_{\rm OO}$ increases from 750~K to $\sim$ 1500~K with decreasing 
ionic radius IR (La $\to$ Dy) \cite{ordertodisorder,dabrowski,Zhou2006,maris}; under increasing pressure
 eventually orbital order melts \cite{Loa,Chen}, while JT distortions still persist in 
 nanoclusters \cite{persistence}.
 
The strength of super-exchange is directly linked to the amplitude of the hopping integrals,  which
depend on the cell volume and distortions. In the REMnO$_3$ series the volume decreases with ionic radius.
Tilting and rotation, however, increase, because of the increasing mismatch between the Mn-O and RE-O bond-lengths. 
For LaMnO$_3$ a volume collapse at $T_{\rm OO}$ has been reported \cite{volume}. Under pressure, up to  P=18 GPa
the volume decreases by $\sim 10\%$, while tilting/rotation slightly decrease. 
A sizable volume reduction typically increases the Mn-O hopping integrals, while tilting and rotation tends to
reduce them, reducing super-exchange effects. The scenario is further complicated by the
local crystal-field  \cite{OO,ruthenates,evad1}, which can, depending on its size and symmetry, help or compete with 
super-exchange, and thus even reverse the trends.  

In this Letter we clarify the role of super-exchange in orbital-melting.
We show that, already in the absence of crystal-field splitting, only in LaMnO$_3$ $T_{\rm KK}\sim T_{\rm OO}$, 
while in all other systems  $T_{\rm KK}$ is 2-3 times smaller than  $T_{\rm OO}$:
While $T_{\rm OO}$ strongly increases with decreasing IR, $T_{\rm KK}$ slightly decreases.
Taking the tetragonal crystal-field splitting into account, these trends are enhanced even further.
This proves that, although  very large, in view of the reported experimental trends,
super-exchange plays a minor role in the orbital-melting transition.
\begin{figure}
\center
\rotatebox {0}{\includegraphics [width=0.4\textwidth]{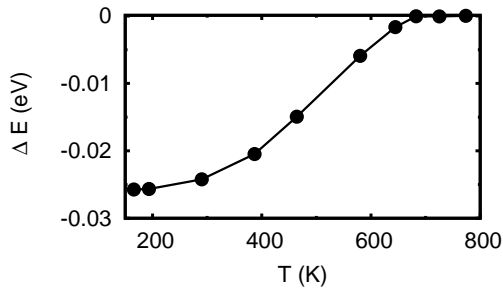}}
\caption{ \label{deltaE}
Energy gain per formula unit due to orbital-ordering (LaMnO$_3$). Error
bars are smaller than the symbols.}
\end{figure}
In order to quantify the role of many-body super-exchange in determining $T_{\rm OO}$,
we perform ab-initio calculations based on the local density approximation (LDA) + dynamical mean-field theory (DMFT) method \cite{lda+dmft}
in the paramagnetic phase. 
The minimal model Hamiltonian to study super-exchange effects  in manganites is the Hubbard model
for the $e_g$ bands in the magnetic field $h=J{S}_{t_{2g}}$ of disordered $t_{2g}$ spins ${\bf S}_{t_{2g}}$\cite{millis}
\begin{eqnarray}\label{ham}
 \nonumber
 H &=&
 \!\!\!\sum_{i} {\varepsilon_{JT}}\tau_{ix}+{\varepsilon_{T}}  \tau_{iz}
  - \!\!\! \!\!\!\!\sum_{i\neq i^\prime m\sigma m'\sigma'} \!\!\! 
   t^{i,i'}_{m,m'} u^{i,i'}_{\sigma,\sigma'} 
   c^{\dagger}_{im\sigma} c^{\phantom{\dagger}}_{i' m'\sigma'}\\
   \nonumber
   &-&h\sum_{im} (n_{im\Uparrow}-n_{im\Downarrow}) 
      +U\sum_{ im }  n_{im\Uparrow }n_{im\Downarrow} \\
   &+&\!\!\frac{1}{2}\!\sum_{im\left( \neq m'\right)\sigma\sigma'}
      \!\!\!(U-2J-J\delta_{\sigma,\sigma'}) n_{ im\sigma} n_{im'\sigma'}\;.
\label{H}
\end{eqnarray}
Here $c_{im\sigma}^{\dagger}$ creates an electron with spin $\sigma\!=\Uparrow,\Downarrow$ in a Wannier orbital $|m\rangle=|x^2-y^2\rangle$ or $|3z^2-1\rangle$  at site $i$, and $n_{im\sigma}=c_{im\sigma}^{\dagger}c^{\phantom{\dagger}}_{im\sigma}$. $\Uparrow$ ($\Downarrow$) indicates the $e_g$ spin parallel (antiparallel) to the ${t_{2g}}$ spins (on that site). In the paramagnetic state, the matrix $u$ ($u^{i,i'}_{\sigma,\sigma'}=2/3$) accounts for the orientational disorder  of the ${t_{2g}}$ spins \cite{millis}; $t^{i,i'}_{m,m'}$ is the LDA \cite{nmto} hopping integral from orbital $m$ on site $i$ to  orbital $m'$ on site $i'\neq i$, obtained {\it ab-initio} by down-folding the LDA bands and constructing a localized $e_g$ Wannier basis. 
The on-site term ${\varepsilon_{JT}}\tau_{ix}+{\varepsilon_{T}}  \tau_{iz}$ yields the LDA
crystal-field matrix. It is the sum of a Jahn-Teller (${\varepsilon_{JT}}\tau_{ix}$) and a tetragonal (${\varepsilon_{T}}\tau_{iz}$) term, where $\tau_{ix}$ and $\tau_{iz}$ are the pseudospin-1/2 operators 
$\tau_{ix}=\frac{1}{2}\sum_{\sigma,m \neq m^\prime}c^\dagger_{im\sigma} c_{im^\prime\sigma}$,
$\tau_{iz}= \frac{1}{2}\sum_{\sigma, m}(-1)^{\delta_{m,x^2-y^2}}c^\dagger_{im\sigma} c_{im\sigma}$.
$U$ and $J$ are the direct and exchange screened on-site Coulomb interaction. We use
the theoretical estimates $J=0.75$~eV, $U\sim5$~eV \cite{MF96,held,OO} and $2JS_{t_{2g}}\sim 2.7$~eV \cite{held}; we find that, in the high-spin regime, $T_{\rm KK}$ is not sensitive to the specific value of $2JS_{t_{2g}}$, therefore we keep $h$ fixed in all results we present.
We solve (\ref{H}) within DMFT \cite{DMFT} using a quantum Monte Carlo \cite{hirsch} solver, working with the full self-energy matrix $\Sigma_{mm'}$ in orbital space \cite{evad1}. %
We construct the LDA  Wannier functions  via the downfolding procedure based on the Nth-Order Muffin-Tin (NMTO) method \cite{nmto}.  Additionally, we perform LAPW calculations \cite{wien2k}, and construct maximally localized Wannier functions \cite{mlwf}. The band-structures and parameter trends obtained with the two methods are very similar \cite{lapw}.

To determine the super-exchange transition temperature T$_{\rm KK}$ we use two independent approaches.
In the first, we calculate the order parameter $p$ as a function of temperature $T$, in the second 
we determine the $T=0$ total energy gain $\Delta E(p)$  (Fig.~\ref{deltaE}) due to orbital order.

\begin{figure}
\center
\rotatebox {0}{\includegraphics [width=0.45\textwidth]{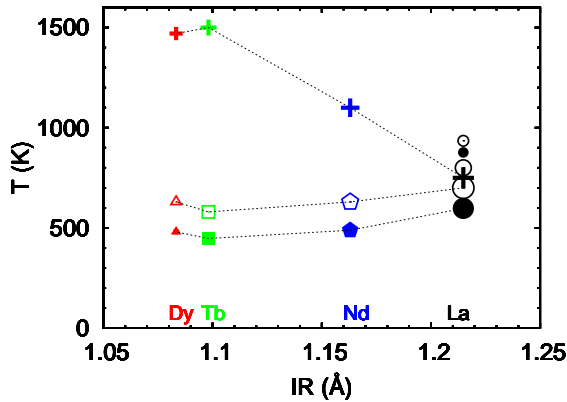}}
\caption{(Color online) \label{TKK}
Orbital-order transition temperature $T_{\rm KK}$ \cite{note} in rare-earth manganites REMnO$_3$ versus RE$^{3+}$ radius,
with RE=Dy (triangles), Tb (squares), Nd (pentagons), La (circles).
Full symbols: $T_{\rm KK}$ from LDA+DMFT total-energy.
Empty symbols: $T_{\rm KK}$ from LDA+DMFT order parameter calculations.
Symbols of decreasing size: P=0 GPa, 5.4 GPa and 9.87 GPa.
Crosses: Experimental values (ambient pressure) from Refs. \cite{Zhou2006,dabrowski,maris}.}
\end{figure}

\begin{figure}
\center
\rotatebox {90}{\includegraphics [width=0.3\textwidth]{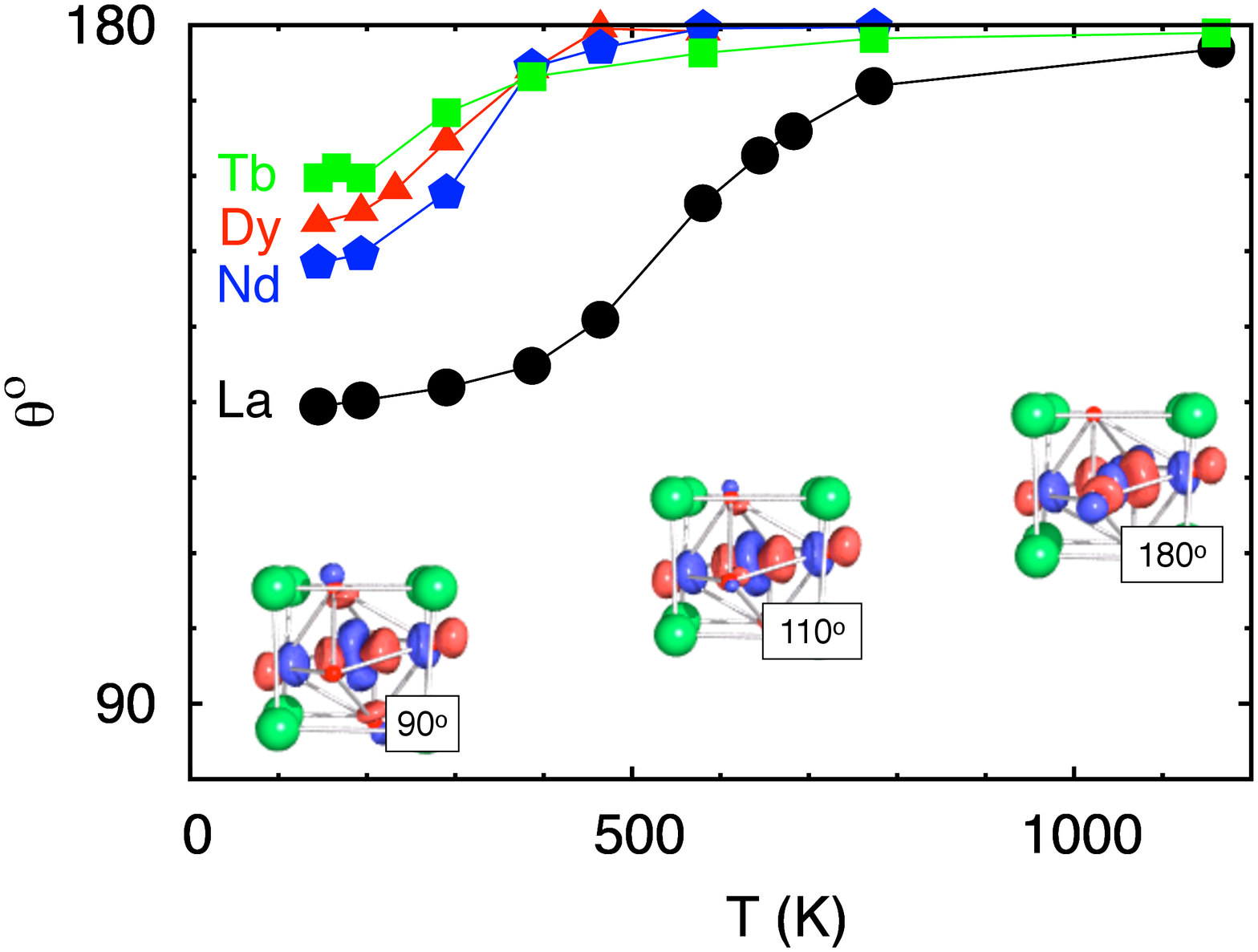}}
\caption{(Color online) \label{tetragonal}
Rotation of the occupied state $|\theta\rangle$ as a function of temperature in
the presence of a 130~meV tetragonal crystal field.
The orbitals are shown for  TbMnO$_3$.}
\end{figure}

The order parameter for orbital-ordering is the orbital polarization $p\equiv |n_1-n_2|$, where $|1\rangle$ and $|2\rangle$
are the natural orbitals in $e_g$-space.  To determine $T_{\rm KK}$ we perform
LDA+DMFT calculations as a function of temperature for all materials in the series. 
They differ in (i) hopping integrals and (ii) crystal field, due to static distortions.  
In order to separate the effects of super-exchange from those of the crystal field, we perform LDA+DMFT calculations
of the orbital polarization as a function of temperature for the real system ($H^{\rm LDA}$), for ideal structures with the same hopping integrals  
but no crystal-field splitting (Fig.~\ref{TKK}), and for ideal structures with only tetragonal splitting
(Fig.~\ref{tetragonal}). 

In the second approach we calculate the energy gain due to orbital order from the difference 
in total energy between the orbitally polarized and the orbitally disordered states,
in the absence of crystal fields ($\varepsilon_{T}=\varepsilon_{ JT}=0$).
We first perform LDA+DMFT calculations for decreasing temperature
and calculate the total energy per formula unit and polarization $p$, $E_{\rm TOT}(p)$.
Next, we repeat the same procedure, but with the constraint 
$p=0$ ($\Sigma_{1,1}=\Sigma_{2,2}$ and $\Sigma_{1,2}=0$).

The total energy is given by \cite{tote},
$$
E_{\rm TOT}(p) = E_{\rm TOT}^\textrm{LDA} + \langle {H} \rangle_p - E_{e_g}^{\rm LDA}  - E_\textrm{DC},
$$
where $E_{\rm TOT}^{\rm LDA}$ is the LDA total-energy, $E_{e_g}^\textrm{\rm LDA}$ the thermal average 
of (\ref{ham}) in the non-interacting ($U=0,J=0$) case, $\langle {H} \rangle_p$ the actual thermal average of (\ref{ham}) in DMFT
for polarization $p$, and $E_{DC}$ the double counting correction. 
Of all these terms only  $\langle {H} \rangle_p$ contributes to 
\begin{equation}
\Delta E (p)=E_{\rm TOT}(p)-E_{\rm TOT}(0)=\langle H \rangle_{p}-\langle H \rangle_{p=0}.
\end{equation}
$\langle {H} \rangle_{p}$ can be split into a single-electron contribution (from the first three terms in (\ref{ham})), 
which we calculate as sum on Matsubara frequencies, and a correlation contribution (from the last two terms in (\ref{ham})),
which we obtain from the double-occupancy matrix. Since $-\Delta E_{\rm TOT} (p) \sim 10-50$ meV,
error bars, in particular  the QMC statistical error on
the double-occupancies matrix, have to be controlled to high accuracy \cite{accuracy}. 
The total-energy gain for LaMnO$_3$ is shown in Fig.~\ref{deltaE}. 
We obtain similar behavior for the other systems.  
In the zero-temperature limit, we extrapolate from $\Delta E (p)$
the super-exchange energy gain $\Delta E_{\rm KK}=E_{\rm TOT}(p=1)-E_{\rm TOT}(p=0)$.
Remarkably, we find that the static mean-field \cite{meanfield} relation $T_{\rm KK}\equiv|2\Delta E_{\rm KK}|/  k_{\rm B}$,
which is valid for spin-1/2 Heisenberg-like model \cite{KK} with arbitrary coupling constants, gives transition temperatures close to those obtained
from order-parameter calculations, the difference being a mere small shift.
Our results are shown in Fig.~\ref{TKK}. While $T_{\rm KK}\sim T_{\rm OO}$ in LaMnO$_3$, in all other systems 
$T_{\rm KK}$ is a factor 2-3 smaller than the experimental estimate for $T_{\rm OO}$.
Moreover, $T_{\rm KK}$ is maximum in LaMnO$_3$, and roughly decreases with IR
from RE=La to Tb, then increases again. $T_{\rm KK}$ also increases under pressure.
These trends are opposite to those reported experimentally for the orbital melting temperature.
They can be ascribed to the increasing distortions along the REMnO$_3$ series, 
and the decrease in volume and tilting/rotation with increasing pressure. 
Finally, for all systems super-exchange favors the occupation of the orbital (signs are given for the site displayed in Fig.~\ref{tetragonal})
$|\theta\rangle= -\sin \frac{\theta}{2}| x^2-y^2\rangle + \cos \frac{\theta}{2}| 3z^2-1 \rangle $, with $\theta=90^o$,
while experimentally $\theta\sim 108^o$ in LaMnO$_3$ increasing with decreasing IR to $114^o$ in TbMnO$_3$ \cite{angles}.

\begin{figure}
\center
\rotatebox {0}{\includegraphics [width=0.5\textwidth]{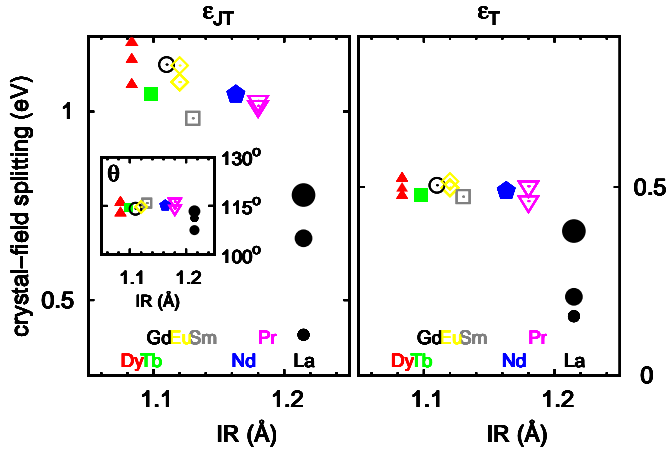}}
\caption{(Color online) \label{cef}
Evolution of the crystal-field (Jahn-Teller and tetragonal) as a function of the rare-earth ion radius
calculated for all structures reported in Refs.~\cite{dabrowski,alonso,mori,esko}. 
Filled circles of decreasing size: LaMnO$_3$ for P=0, 5.4 and 9.87 GPa \cite{Loa}. Inset: calculated occupied orbital.
}
\end{figure}

Due to the competition between the tetragonal crystal-field splitting $\varepsilon_{T}$ 
and super-exchange (which favor the occupation of different orbitals), 
$T_{\rm KK}$ is reduced even further. We find that for finite $\varepsilon_T$ the system is orbitally ordered 
already at high temperature due to the crystal field, but the occupied orbital has $\theta=180^0$.
In Fig.~\ref{tetragonal} we show the results for $\varepsilon_T$ fixed at $\sim 130$~meV,
sizable but smaller than for any of the considered systems (see Fig.~\ref{cef}).
We find that at the reduced critical temperature $T_{\rm KK}^{\varepsilon_{T}}$, super-exchange rotates 
the orbital towards $90^o$.
The change in $T_{\rm KK}$ is small for LaMnO$_3$, but $T_{\rm KK}$ is reduced to 400~K for NdMnO$_3$, and even
more for DyMnO$_3$ and TbMnO$_3$. 
Furthermore, in the zero-temperature limit, the smaller $T_{\rm KK}^{\varepsilon_{T}}$, the closer is
$\theta$ to $180^o$.  Thus a fixed $\varepsilon_T\sim 130$~meV enhances the trend found for $\varepsilon_T=0$:
$T_{\rm KK}$ is larger in LaMnO$_3$, and decreases going to DyMnO$_3$. Still, even for LaMnO$_3$,
$\theta$ is significantly larger than the experimental $108^o$. This means that a
Jahn-Teller crystal-field splitting $\varepsilon_{ JT}$ is necessary to explain the experimental $\theta$;
Fig.~\ref{tetragonal} shows that such splitting has to increase for the series  RE=La, Nd, Dy, Tb. 
Taking into account that tetragonal splitting actually increases with decreasing pressure,
and substituting La with Nd, Tb, or Dy (Fig.~\ref{cef}), this trend is enhanced even more.
For  $\varepsilon_T$ corresponding to the real structures, down to 150~K
we find no super-exchange transition  for all systems but LaMnO$_3$.
These results can be understood qualitatively in static mean-field theory. In this approach,
super-exchange yields an effective  Jahn-Teller splitting 
$\varepsilon_{\rm KK}=\langle \tau_x\rangle \lambda_{\rm KK}$,  where $\lambda_{\rm KK}$ is 
the molecular field parameter; the self-consistency condition
for orbital order is
$\langle \tau_x\rangle =\frac{1}{2}\sin \theta \tanh \left(\beta \sqrt{\varepsilon_{ T}^2 + \varepsilon_{\rm KK}^2}/2 \right)$,
with $\sin\theta=  \varepsilon_{\rm KK}/\sqrt{\varepsilon_{ T}^2 + \varepsilon_{\rm KK}^2}$.
This equation has a non-trivial solution ($\theta\neq180^o$) only if $\lambda_{\rm KK}/2>\varepsilon_T$. The critical
temperature is $T_{\rm KK}^{\varepsilon_{ T}}/T_{\rm KK}^0=( \varepsilon_T/2k_BT_{\rm KK}^0)/\tanh^{-1} (\varepsilon_T/2k_BT_{\rm KK}^0)$,
with  $k_BT_{\rm KK}^0=\lambda_{\rm KK}/4$; it decreases with increasing $\varepsilon_T$, while $\theta \to 180^o$ \cite{notedelta}.
For large enough $\varepsilon_T$ ({$\varepsilon_T > \lambda_{\rm KK}/2$}) there is no super-exchange driven transition at all.

In conclusion, for the orbital-melting transition in rare-earth manganites REMnO$_3$, we
find that many-body super-exchange yields a transition temperature $T_{\rm KK}$ very close to T$_{\rm OO}$ only  in 
LaMnO$_3$, while in all other systems $T_{\rm KK}$ is less than half $T_{\rm OO}$. 
Moreover, we find that super-exchange yields $\theta \sim 90^o$
for the occupied orbital, while in the experimental structures $\theta\sim 108^o-114^o$.
We also find that a tetragonal splitting $\epsilon_T$ reduces $T_{\rm KK}$ even further. $\epsilon_T$
increases substituting La with Nd, Tb or Dy and decreases under pressure. Finally,
super-exchange effects become larger with increasing pressure, while experimentally
orbital order eventually melts \cite{Loa,Chen}.
Our work proves that, in the light of the experimentally observed trends, super-exchange plays a minor role 
in the orbital-melting transitions of rare-earth manganites. 

We thank I.~Loa and K.~Syassen for sharing unpublished data.
Calculations were done on the J\"ulich Blue Gene/P. We acknowledge financial support from 
the  Deutsche Forschungsgemeinschaft through research unit FOR1346.


\begin{thebibliography}{99}
\bibitem{accuracy} A simple estimate from the average correlation energy, $\frac{1}{2} U n_{e_g}(n_{e_g}-1)$,
is $\frac{1}{2} (U (2n_{\rm e_g}-1)) \delta  n_{\rm e_g}\ll |\Delta E (p)|$,
which for $\Delta E (p)\sim-10$~meV, $n_{\rm e_g}=1$ and $U=5$~eV,  yields $\delta n_{e_g}\ll 4\cdot10^{-3}$.
\bibitem{note}  
All results shown are for $U=5$~eV and $J=0.75$~eV. For $P=9.87$~GPa, setting
$\varepsilon_{ T}=\varepsilon_{ JT}=0$ we obtain however a metallic solution with $p=0$.
Since $T_{\rm KK}$ decreases with $U$ roughly as $\sim 1/U$,  
as expected from super-exchange theory, to compare values for constant $U$, 
we then extrapolate the $U=5$~eV value of $T_{\rm KK}$ from the insulating state
obtained for slightly larger $U$ ($U=5.5$~eV, $U=6$~eV).
\bibitem{notedelta} The energy gain due to the pseudospin rotation is $2|\Delta E |/k_BT_{\rm KK}^{\varepsilon_{ T}}=(1-x)^2(\tanh^{-1} x)/x$, where
$x=2\varepsilon_{ T}/\lambda_{\rm KK}$. For $\theta=130^o$ (as for LaMnO$_3$, Fig.~\ref{tetragonal}) 
we find $T_{\rm KK}^{\varepsilon_{ T}}\sim 0.84 T_{\rm KK}^0$ 
and $2|\Delta E|/k_BT_{\rm KK}^{\varepsilon_{ T}}\sim 0.15$. This ratio is in qualitative
agreement with what we find with DMFT for LaMnO$_3$, $|\Delta E|\sim 10$~meV.
\end{thebibliography}
\end{document}